\documentclass[%
prl,
reprint,nodate,
superscriptaddress,
amsmath,amssymb,
aps,
]{revtex4-1}

\pdfoutput=1

\usepackage[utf8]{inputenc} 

\usepackage{xcolor}

\usepackage{graphicx}
\usepackage{dcolumn}
\usepackage{bm}

\usepackage{hyperref} 
\newcommand{\rf}[1]{Eq.~(\ref{#1})}

\newcommand{\mysection}[1]{{\vspace{10 pt}\noindent \emph{{\textbf{#1}}--} }}

\newcommand{\f}[2]{\frac{#1}{#2}}

\newcommand{\sym}{${\mathcal N}=4$}

\makeatletter
\newcommand*{\balancecolsandclearpage}{%
	\close@column@grid
	\clearpage
	\twocolumngrid
}
\makeatother

\begin{document}
	
	\title{Hydrodynamic attractors in phase space}

	\author{Michal P.\ Heller} \email{michal.p.heller@aei.mpg.de}
	\affiliation{Max Planck Institute for Gravitational Physics, Potsdam-Golm, 14476, Germany}
	
	\affiliation{National Centre for Nuclear Research, 02-093 Warsaw, Poland}
	
	\author{Ro Jefferson} \email{rjefferson@aei.mpg.de}
	\affiliation{Max Planck Institute for Gravitational Physics, Potsdam-Golm, 14476, Germany}
	
	\author{Micha\l\ Spali\'nski}
	\email{michal.spalinski@ncbj.gov.pl}
	\affiliation{National Centre for Nuclear Research, 02-093 Warsaw, Poland}
	
	\affiliation{Physics Department, University of Bia{\l}ystok,
		15-245 Bia\l ystok, Poland}
	
	\author{Viktor Svensson}
	\email{viktor.svensson@aei.mpg.de}
	\affiliation{National Centre for Nuclear Research, 02-093 Warsaw, Poland}
	\affiliation{Max Planck Institute for Gravitational Physics, Potsdam-Golm, 14476, Germany}

	\date{\today}
	
	\begin{abstract}
		
		Hydrodynamic attractors have recently gained prominence in the context of early stages of ultra-relativistic heavy-ion collisions at the RHIC and LHC. We critically examine the existing ideas on this subject from a phase space point of view. In this picture the hydrodynamic attractor can be seen as a special case of the more general phenomenon of dynamical dimensionality reduction of phase space regions. We quantify this using Principal Component Analysis. Furthermore, we adapt the well known slow-roll approximation to this setting. These techniques generalize easily to higher dimensional phase spaces, which we illustrate by a preliminary analysis of a dataset describing the evolution of a 5-dimensional manifold of initial conditions immersed in a 16-dimensional representation of the phase space of the Boltzmann kinetic equation in the relaxation time approximation. 
		
	\end{abstract}
	
	\maketitle
	
	\mysection{Introduction} The physics of strong interactions studied in heavy-ion collisions at the RHIC and LHC (see e.g. Ref.~\cite{Busza:2018rrf} for a concise contemporary review) has been a remarkable source of inspiration for the study of complex systems far from equilibrium. The phenomenological success of relativistic hydrodynamics, together with calculations in microscopic models based on holography and kinetic theory, have inspired several novel research directions. One such direction is centered on the notion of hydrodynamic attractors. These were introduced in Ref.~\cite{Heller:2015dha} with the aim of capturing universal features of non-equilibrium physics beyond the limitations of the gradient expansion and were subsequently explored in many works, including Refs.~\cite{Basar:2015ava,Aniceto:2015mto,Romatschke:2017vte,Spalinski:2017mel,Strickland:2017kux,Romatschke:2017acs,Denicol:2017lxn,Florkowski:2017jnz,Behtash:2017wqg,Casalderrey-Solana:2017zyh,Blaizot:2017ucy,Almaalol:2018ynx,Denicol:2018pak,Behtash:2018moe,Spalinski:2018mqg,Strickland:2018ayk,Strickland:2019hff,Blaizot:2019scw,Jaiswal:2019cju,Strickland:2019hff,Kurkela:2019set,Denicol:2019lio,Brewer:2019oha,Behtash:2019qtk,Chattopadhyay:2019jqj,Dash:2020zqx,Shokri:2020cxa,Almaalol:2020rnu,Blaizot:2020gql,Mitra:2020mei,mazeliauskas2019prescaling}.
	
	In the context of reproducing the spectra of soft particles in ultra-relativistic heavy-ion collisions, the underlying observable of interest is the expectation value of the energy-momentum tensor $\langle T^{\mu \nu} \rangle$. Ideally, one would like to have a way of predicting its behaviour as a function of time directly in QCD. Such calculations remain beyond reach, but have been pursued in quantum field theories possessing a gravity dual~\cite{Maldacena:1997re, Witten:1998qj, Gubser:1998bc} in a number of works, see Refs.~\cite{Chesler:2015lsa,Berges:2020fwq} for a review. Another line of development replaces QCD by its effective kinetic theory description~\cite{Arnold:2002zm}, as reviewed in Refs.~\cite{Schlichting:2019abc,Berges:2020fwq}.
	The evolution of $\langle T^{\mu \nu} \rangle$ depends on the initial state, captured by the bulk metric in holography, or by the initial distribution function in the kinetic theory. After some time, the vast majority of this information is effectively lost, and $\langle T^{\mu \nu} \rangle$ evolves hydrodynamically. These explorations have led to the idea of a hydrodynamic attractor, identified in Ref.~\cite{Heller:2015dha} in a class of hydrodynamic theories~\cite{Muller:1967zza, Israel:1979wp, Baier:2007ix} as a specific solution which is approached by generic solutions initialized at arbitrarily small times.
	It was also observed there that a ``slow-roll'' condition akin to what is used in inflationary cosmology~\cite{Liddle:1994dx, Spalinski:2007kt} leads to an accurate approximation of this attractor. Subsequent works have lead to many  interesting phenomenological applications~\cite{Heller:2015dha, Romatschke:2017vte, Behtash:2018moe, Giacalone:2019ldn, Dore:2020fiq}.
	
	Despite these developments, there are three important yet largely unexplored issues. The first addresses the different concepts of attractor, the second concerns their relevance for the dynamics of initial states of interest and the third is the question of their existence beyond highly-symmetric settings. We address these points with the goal of clearing the ground for new developments, in particular for generalisations to more realistic flows with less symmetry. Our approach is based on the phase space picture, i.e., the space of variables needed to parametrize the dynamics underlying $\langle T^{\mu \nu} \rangle$, which was introduced in this context in Refs.~\cite{Behtash:2017wqg, behtash2019global}.
	
	\mysection{Dissipation of initial state information} There are two key features of the dynamics following an ultra-relativistic heavy-ion collision: the expansion which drives the system away from equilibrium and interactions which favour equilibration~\cite{Blaizot:2017ucy}. The simplest model of this is Bjorken flow which assumes one-dimensional expansion and boost-invariance along the expansion axis, conveniently described in terms of proper time $\tau$ and spacetime rapidity $y$. In interacting conformal theories, at asymptotically late times the system follows a scaling solution for the (effective) temperature~\cite{Bjorken:1982qr}
	\begin{equation} \label{bjorken}
	T(\tau) = \f{\Lambda}{(\Lambda\tau)^{1/3}} + \dots~.
	\end{equation}
	In this equation, the dimensionful constant $\Lambda$ is the only trace of initial conditions. Corrections to this come in two forms: higher-order power-law terms 
	which are sensitive only to $\Lambda$, and exponential corrections which encode information about the initial conditions and describe its dissipation ~\cite{Janik:2006gp, Heller:2013fn, Heller:2015dha, Casalderrey-Solana:2017zyh, Aniceto:2015mto, Aniceto:2018uik}. This asymptotic form is usually referred to as a transseries~\cite{Aniceto:2018bis}. The simplest models where this can be observed explicitly are formulated in the language of hydrodynamics.
	
	\mysection{Models of hydrodynamics} We primarily focus on hydrodynamic theories (see Ref.~\cite{Florkowski:2017olj} for a review), which despite their name include transient non-hydrodynamic excitations needed to avoid acausality. In such models, $\langle T^{\mu \nu} \rangle$ is decomposed into a perfect fluid term and a ``dissipative'' part denoted~by~$\pi^{\mu \nu}$:
	\begin{equation}\label{defPimunu}
	\langle T^{\mu \nu} \rangle = ({\cal E} + {\cal P}) \, u^{\mu} u^{\nu} + {\cal P}\, g^{\mu \nu} + \pi^{\mu \nu},
	\end{equation}
	where $u_{\mu} u^{\mu} = -1$, $u_{\mu} \pi^{\mu \nu} = 0$, and the energy density $\cal E$ and pressure $\cal P$ are related by the thermodynamic equation of state. In the conformal case considered here, ${\cal P} = \frac{\cal E}{3}$. Conservation equations of $\langle T^{\mu \nu} \rangle$ provide the four equations of motion for $\cal E$ and $u^{\mu}$. Hydrodynamic models, building on the original ideas of Refs.~\cite{Muller:1967zza, Israel:1979wp}, provide the remaining equations for $\pi^{\mu \nu}$ in terms of relaxation-type dynamics that ensure matching to the hydrodynamic gradient expansion of any microscopic model. 
	
	In this Letter, we consider two classes of models. The first one is the M{\"u}ller-Israel-Stewart (MIS) model~\cite{Muller:1967zza,Israel:1979wp} 
	\begin{equation}\label{eq.BRSSS}
	\tau_{\pi} \, {\cal D} \, \pi^{\mu \nu} = - \pi^{\mu \nu} + \eta \, \sigma^{\mu \nu},
	\end{equation}
	where {${\cal D} \, \pi^{\mu \nu} \equiv u^{\alpha} \nabla_{\alpha} \pi^{\mu \nu} + \ldots$} is the Weyl-covariant derivative in the co-moving direction \cite{Loganayagam:2008is, Heller:2014wfa} and $\sigma^{\mu \nu} = 2\,{\cal D}^{(\mu} u^{\nu)}$ is the shear tensor.
	One can supplement Eq.~\eqref{eq.BRSSS} with additional terms defining the so-called Baier-Romatschke-Son-Starinets-Stephanov (BRSSS) model~\cite{Baier:2007ix} and in the following we will refer to it as MIS/BRSSS. 
	Conformal symmetry requires that
	\begin{equation}\label{eq.defsBRSSS}
	\eta = C_{\eta} \frac{{\cal E} + {\cal P}}{T} \quad \mathrm{and} \quad \tau_{\pi} = \frac{C_{\tau_{\pi}}}{T} \quad \mathrm{with} \quad {\cal E} \sim T^{4},
	\end{equation}
	where $C_{\eta}$, $C_{\tau_{\pi}}$ are dimensionless constants and $T$ is defined as the temperature of an equilibrium state at the same energy density~$\cal E$. Eq.~\eqref{eq.BRSSS} implies relaxation phenomena on a time scale defined by the relaxation time~$\tau_{\pi}$.
	
	For the Bjorken flow, the combined equations reduce to a second order ordinary differential equation (ODE) for the effective temperature $T(\tau)$, see Eq.~(7.17) in Ref.~\cite{Florkowski:2017olj}. 
	The hydrodynamic attractor was originally observed in this model in Ref.~\cite{Heller:2015dha} using a special scale-invariant parametrization involving pressure anisotropy (note~$\pi^{2}_{\, 2} = \pi^{3}_{\, 3}$)
	\begin{equation}\label{pa}
	{\cal A} = \frac{\pi^{2}_{\, 2} - \pi^{y}_{\, y}}{{\cal P}} = 6 + 18 \, \tau \, \f{\dot{T}(\tau)}{T(\tau)}
	\end{equation}
	understood as a function of time measured by 
	\begin{equation}\label{def.w}
	w = \tau \, T(\tau)~.
	\end{equation}
	We denote derivatives with respect to~$\tau$ with a dot and derivatives with respect to~$w$ with a prime. In contrast with $T(\tau)$, ${\cal A}(w)$ satisfies a first order ODE, see Eq.~(7.18) in Ref.~\cite{Florkowski:2017olj}.
	
	The second hydrodynamic theory of interest here is the Heller-Janik-Spalinski-Witaszczyk (HJSW) model~\cite{Heller:2014wfa} (see also Ref.~\cite{Noronha:2011fi}), in which \rf{eq.BRSSS} is replaced by
	\small
	\begin{equation} \label{eq.HJSW}
	\big\{ (\frac{1}{T}{\cal D})^2 + \frac{2}{T^2 \tau_{\pi}}{\cal D} \big\} \pi^{\mu \nu} = - \f{\tau_{\pi}^{-2} + \omega^2}{T^2} \big\{ \pi^{\mu \nu} +  \eta\sigma^{\mu \nu} \big\}.
	\end{equation}
	\normalsize
	This structure again ensures relaxation phenomena on a time scale~$\tau_{\pi}$, but here  they occur in an oscillatory manner as in \sym\ supersymmetric Yang-Mills theory~\cite{Kovtun:2005ev}, with frequency~$\omega = C_{\omega} T$. This model leads to a third order ODE for the effective temperature $T(\tau)$ or a second order ODE for ${\cal A}(w)$, see Eq.~(7.26) in Ref.~\cite{Florkowski:2017olj}, and provides a workable setting with a richer set of initial conditions than offered by MIS/BRSSS. 
	
	\mysection{Hydrodynamic attractors} The hydrodynamic attractor in MIS/BRSSS arose through studying a range of solutions for~${\cal A}(w)$ 
	which converge and then evolve toward local thermal equilibrium~\cite{Heller:2015dha}. This behaviour is known in the mathematical literature~\cite{kloeden2011nonautonomous} as a \emph{forward attractor}. Intuitively, forward attractors are solutions that attract nearby sets as $w \rightarrow \infty$. In MIS/BRSSS every solution is a forward attractor. Conversely, by considering solutions initialized at earlier and earlier times 
	we observe that generic solutions (which diverge at $w=0$) decay to a specific solution which is regular there. Such behaviour is known as a \emph{pullback attractor}. 
	
	These two notions of attractors, introduced in this context in~\cite{behtash2019global} are concerned with different regimes: the forward attractor describes asymptotically late times while the pullback attractor is a statement about early times. In MIS/BRSSS there is a unique solution which is both a pullback attractor and a forward attractor; we denote this solution by~${\cal A}_\star(w)$. 
	
	\mysection{Attractors and phase space} 
	The above discussion raises two concerns. 
	The first is that were one to visualize a range of solutions 
	$T(\tau)$, the hydrodynamic attractor would not be apparent, and yet it is encoded there since $\mathcal{A}(w)$ can be clearly derived from $T(\tau)$. 
	If a different observable showed attractor behaviour  
	how would one find it? This issue must be clarified if one aims to identify attractors in more complicated settings. We address it by considering the full \emph{phase space} parameterised by \mbox{$(\tau, T, \dot{T})$} for MIS/BRSSS and \mbox{$(\tau, T, \dot{T},\ddot{T})$} for HJSW.
	We include proper time as one of the phase space variables because we are dealing with a non-autonomous system.
	
\begin{figure}[t!]
		\center
		\includegraphics[width=.75\columnwidth]{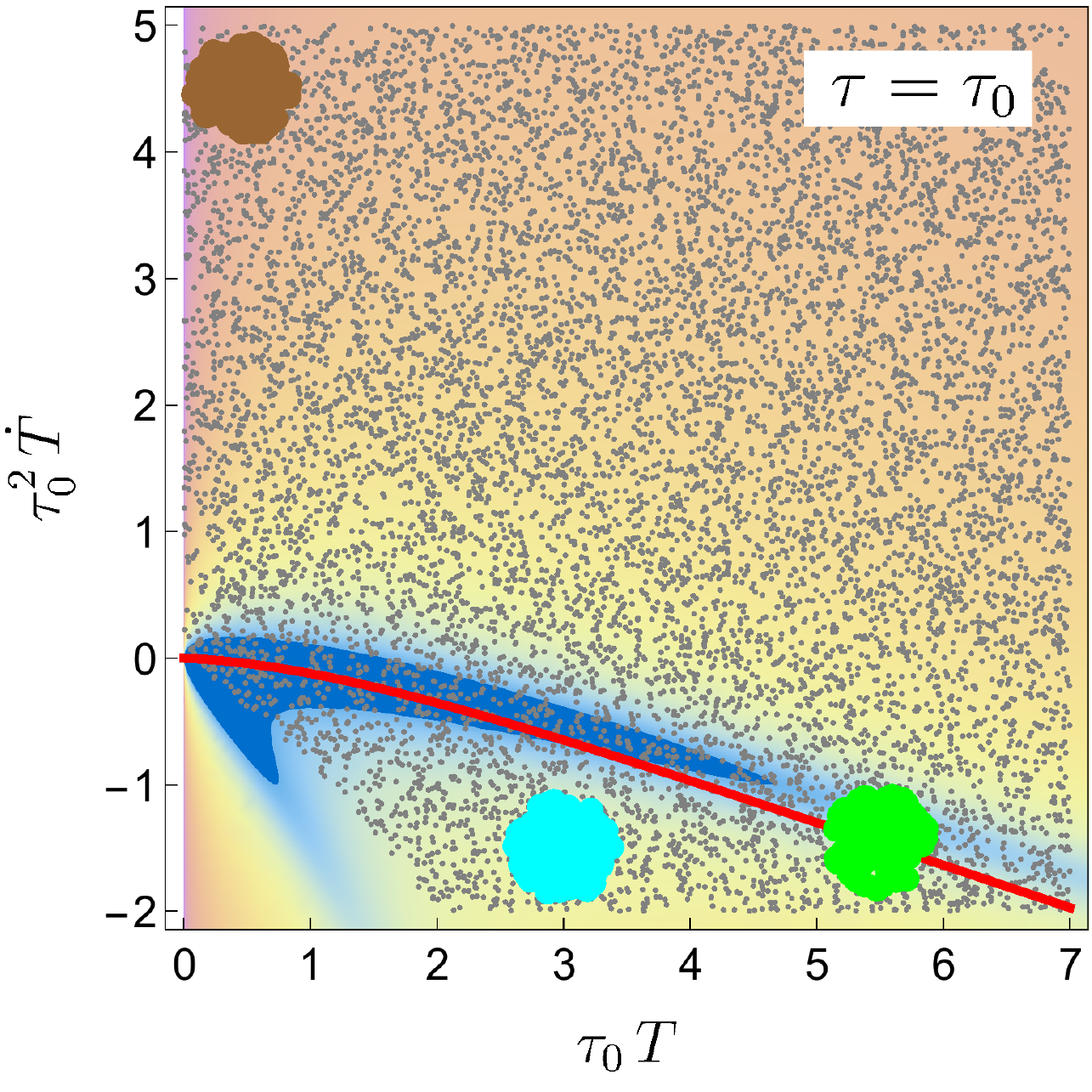}
		\vspace{1 pt}
		\includegraphics[width=.75\columnwidth]{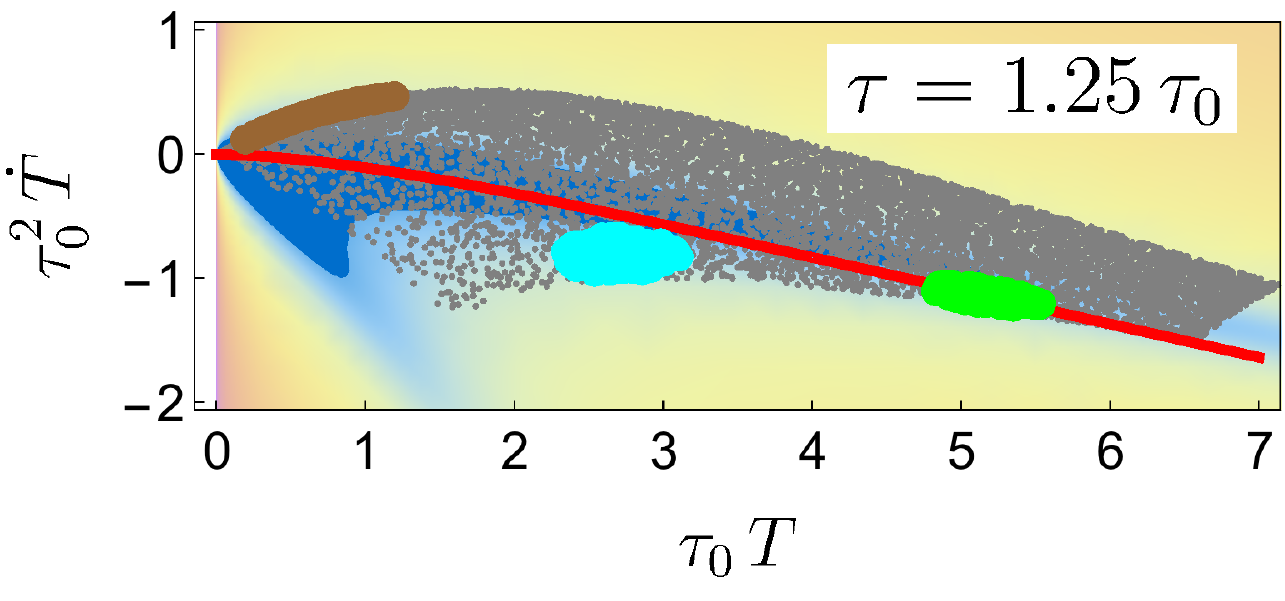}
		\vspace{1 pt}
		\includegraphics[width=.75\columnwidth]{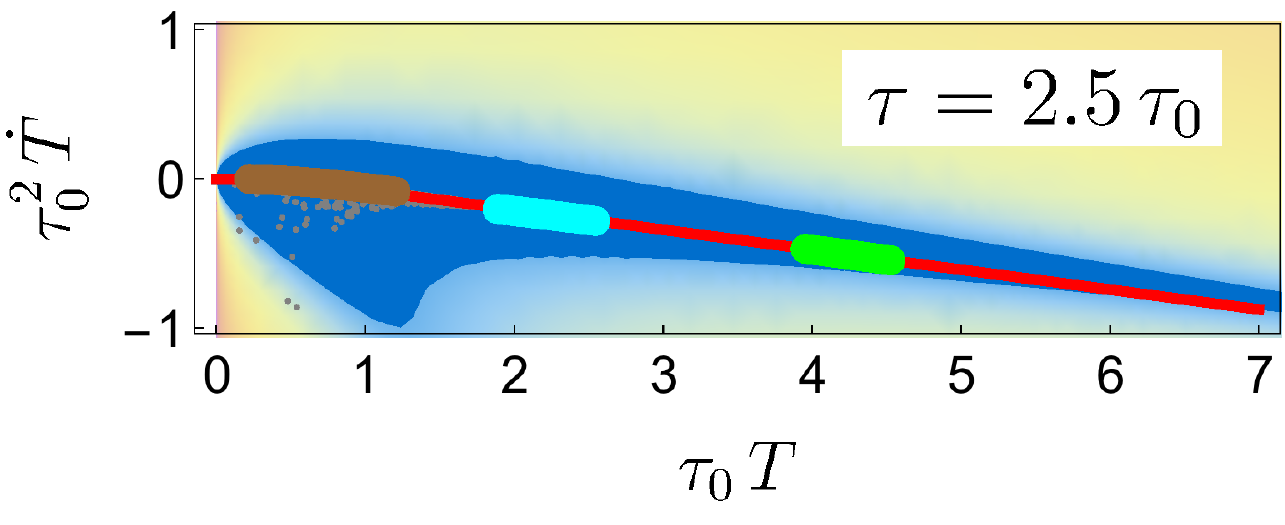}
		\vspace{1 pt}
		\includegraphics[width=.75\columnwidth]{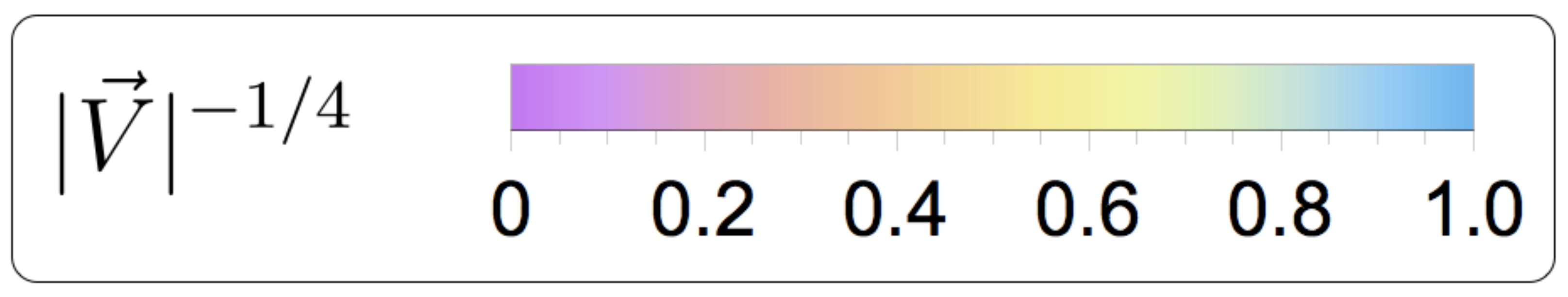}
		\vspace{-5 pt}
		\caption{Three snapshots of evolution in MIS/BRSSS
			phase space of a cloud of about 10,000 random states in the region  determined by the ranges of the top plot. The red curve denotes the family of solutions corresponding to the attractor ${\cal A}_\star(\tau \, T)$. 
			The background color represents the speed 
			at which the points move in phase space, 
			with magenta being faster than blue according to the velocity defined in the text. 
			The dark blue denotes the slow region beyond the color coding scale. For the purposes of visualizing local dimensionality reduction, see also Fig.~\ref{fig:brsss_pca}, we track three initially spherical regions. The plots were made for $C_{\eta} = 0.75$ and $C_{\tau_{\pi}} = 1$, see Eqns.~\eqref{eq.BRSSS} and~\eqref{eq.defsBRSSS}, and $\tau_{0}$ denotes initialization~time.}
		\label{plot:descent}
	\end{figure}

	The second concern  stems from the fact that forward and pullback attractors strongly depend on either asymptotically late or asymptotically early time dynamics. Such asymptotic regimes may be inaccessible or unphysical. We address this by focusing on the \emph{local dynamics} on families of constant-$\tau$ slices of phase space, on which solutions of the equations of motion appear as points (see Figs.~\ref{plot:descent} and~\ref{plot:HJSW}). However, any particular solution ${\cal A}(w)$ corresponds to a different curve on each slice, 
	as dictated by \rf{pa}. On the basis of earlier studies one  expects that as $\tau$ increases, different solutions (points on constant-$\tau$ slices) will collapse to 
	the curves representing the attractor ${\cal A}_{*}(w)$. 
	This is what one \emph{eventually} sees in Figs.~\ref{plot:descent}~(bottom) and~\ref{plot:HJSW}~(bottom-right), but numerical studies of phase space histories reveal a much finer picture. The process of information loss occurs in three phases: local dimensionality reduction, approach to the hydrodynamic attractor loci (red curves in Fig.~\ref{plot:descent}), and evolution toward equilibrium along the attractor. Consider a finite ``cloud'' of initial states, such as any one of the three colored sets of points shown in Fig.~\ref{plot:descent}. Each cloud contracts and becomes one-dimensional. When this happens depends on the initial conditions. For example, the brown cloud in Fig.~\ref{plot:descent} (the one initialized at smallest value of $\tau_0 T$) loses a dimension quite early and rather far from the attractor, while the other two do this much later.

	\mysection{Quantifying dimensionality reduction} 
	We have argued that dimensionality reduction in phase space is an important feature of hydrodynamic attractors, but so far we have not provided a working recipe to quantify this process. A promising direction, which we only begin to explore here, follows from recognizing that dimensionality reduction is one of the basic tasks of machine learning. For the simple cases considered here, Principal Component Analysis (PCA) is quite effective
	(see Refs.~\cite{Bhalerao:2014mua, Mazeliauskas:2015efa,Bozek:2017thv, Liu:2019jxg, Gardim:2019iah} for other applications of PCA to problems in ultra-relativistic heavy-ion collisions). Intuitively, PCA quantifies the variations of a data set in different directions
	and associates an \textit{explained variance} with each of them. 
	
\begin{figure}[t]
		\centering
		\includegraphics[width=.75\columnwidth]{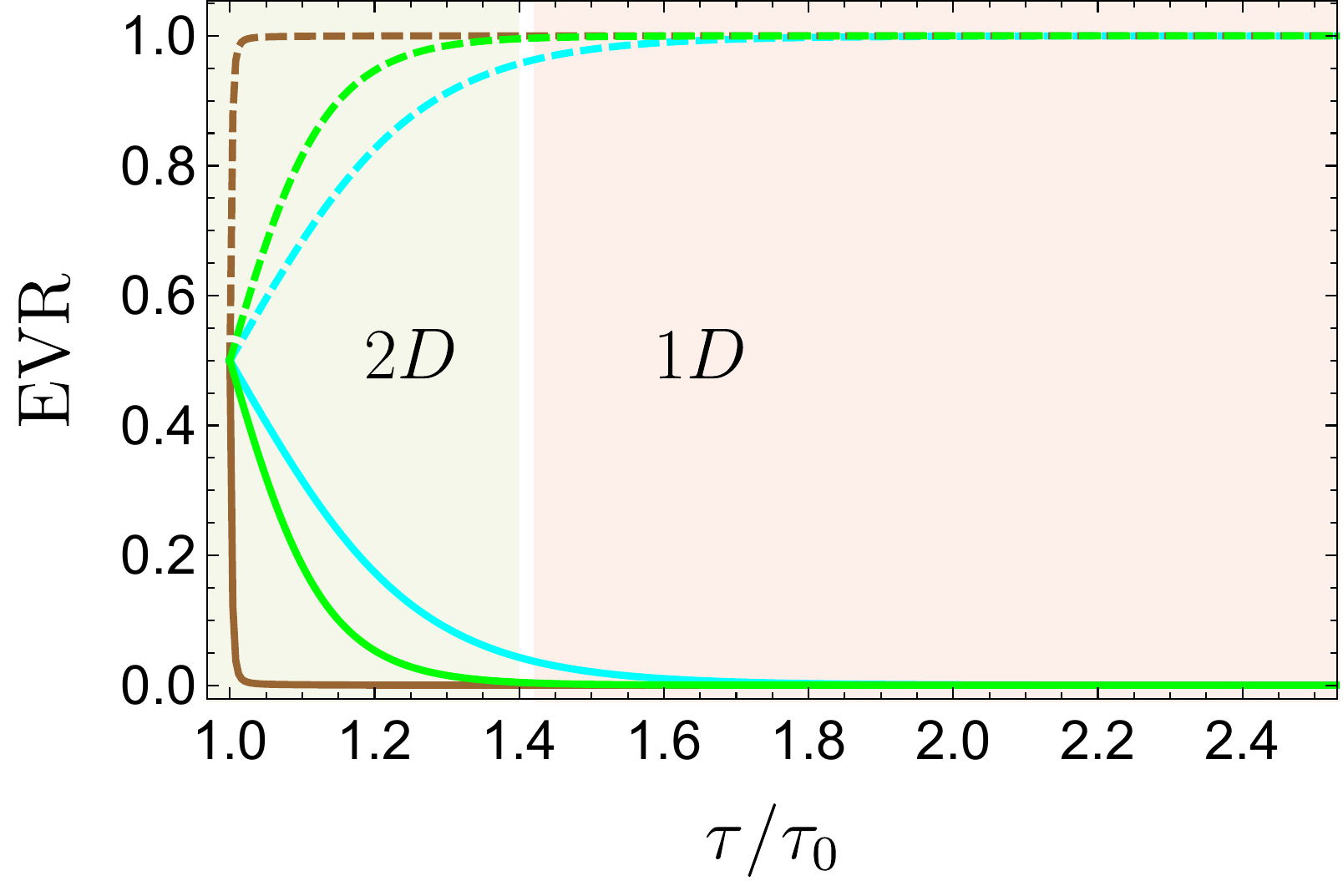} \includegraphics[width=.75\columnwidth]{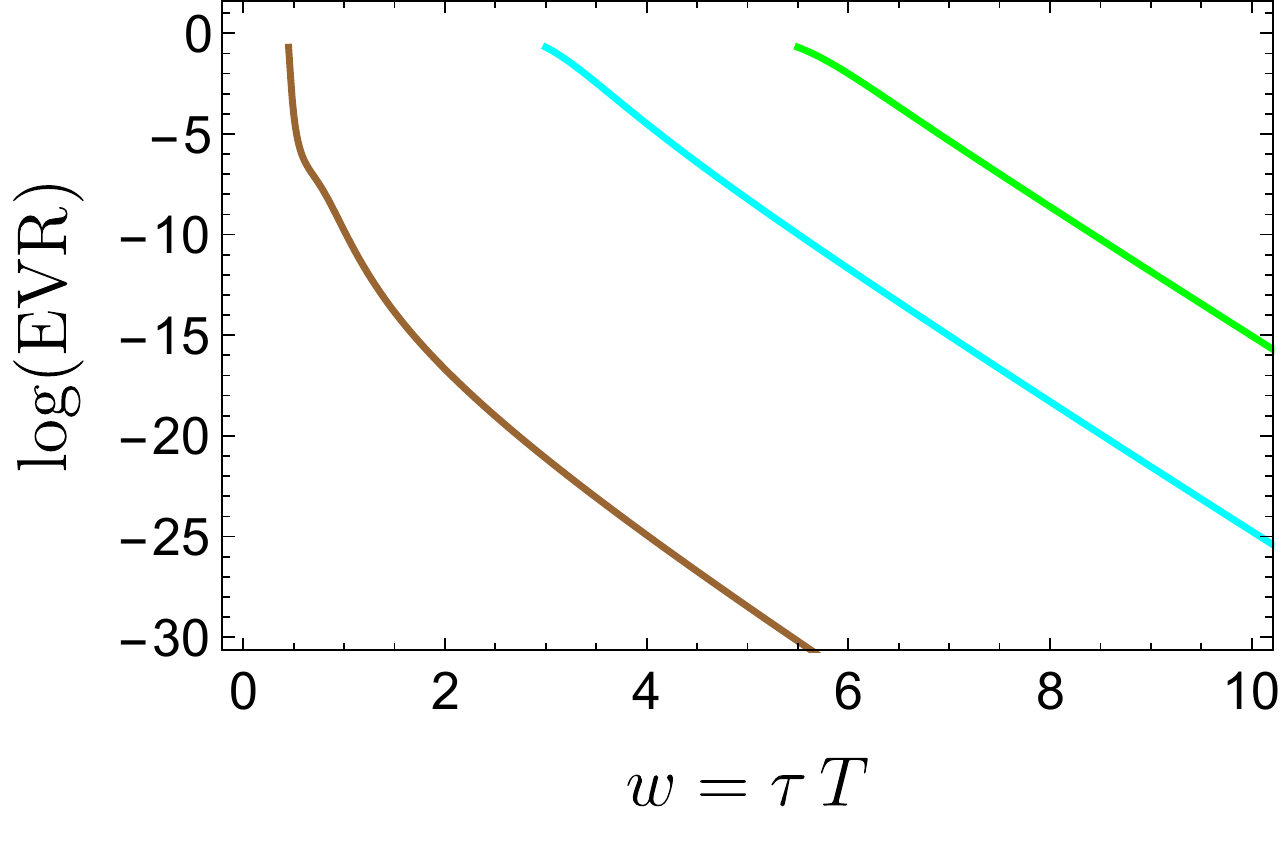}
		\caption{Top: Evolution of explained variance ratio of each principal component in MIS/BRSSS for circles (radius: $10^{-4}$) of initial conditions with centres lying in the middle of initial dots of corresponding color in Fig.~\ref{plot:descent}. Bottom: 
			For large enough values of~$w$ there is an exponential decay
			with a decay rate consistent with 
			twice the transient mode contribution to ${\cal A}(w)$. 
			The initial dimensional reduction of the brown region 
			we view as triggered by the expansion rather than by nonhydrodynamic mode decay~\cite{Kurkela:2019set}.
		}
		\label{fig:brsss_pca}
	\end{figure}

\begin{figure}
		\centering
		\includegraphics[width=1\linewidth]{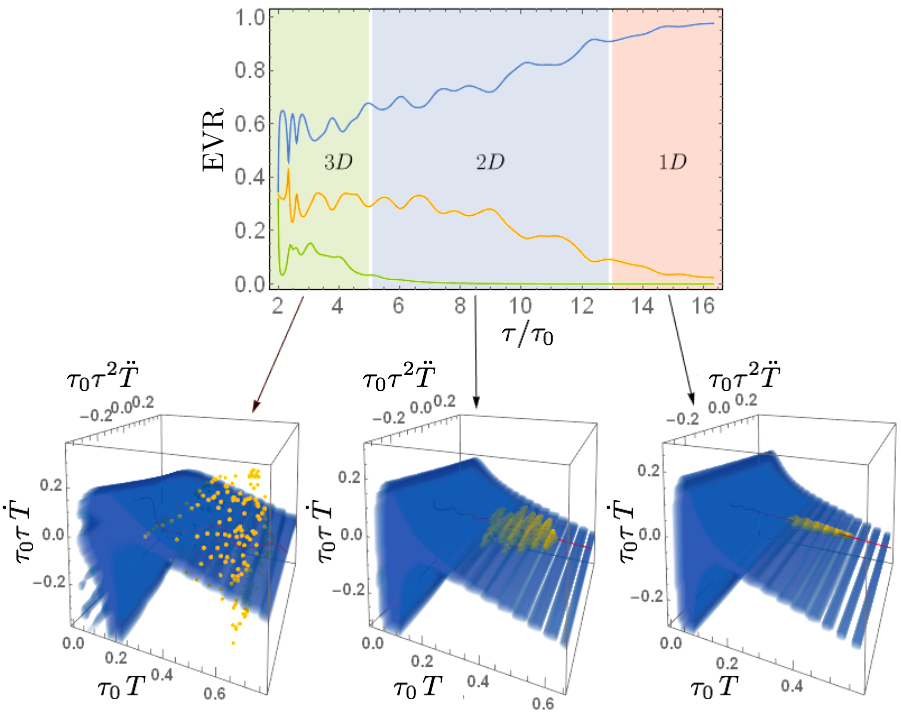}
		\caption{In HJSW, the evolution of a cloud in phase space can be split into three stages, corresponding to the dimensionality of the cloud. 
			The reduction from three to two dimensions corresponds to a collapse onto the slow region (blue region in plots). 
			The plots were made with $C_{\eta} = 0.75$, $C_{\tau_{\pi}} = 1.16$ and $C_{\omega} = 9.8$.}
		\label{plot:HJSW}
	\end{figure}

	We start by applying PCA to the two-dimensional phase space of MIS/BRSSS. On the initial time slice 
	we pick a state $(T,\dot{T})$ and consider a random set of points within a disc around it. For this set of points, the two principal components are approximately equal in magnitude. At each timestep
	we recompute the principal components for the set of states we started with (see Fig.~\ref{plot:descent}) and their evolution in time is shown in Fig.~\ref{fig:brsss_pca}. 
	Dimensionality reduction is signalled when one of the components is much smaller than the other one.
	
	This analysis extends to phase spaces of arbitrary dimension. An interesting example is provided by the three-dimensional phase space of HJSW. The evolution of principal components is shown in the upper part of Fig.~\ref{plot:HJSW}. There are three stages of dimensionality reduction.
	The first, from three to two dimensions, is analogous to the earliest phase of the collapse in MIS/BRSSS, most likely due to the expansion. 
	The second stage is characterized by oscillations 
	typical of the non-hydrodynamic sector of HJSW. 
	These eventually dissipate resulting in the final stage of one-dimensional evolution. 
	
	\mysection{Slow-roll in phase space} The basic intuition is that the attractor locus should correspond to a region where the flow in phase space is slowest.
	One way to motivate why the slow region should behave as an attractor is to use an argument inspired by thermodynamics: a system is likely to be found in a large entropy macrostate because such states cover the majority of phase space. In our setting, the system is likely to be in a slow region because it takes a long time to escape it, while the fast regions can be quickly traversed.
	
	We have found that in the case of MIS/BRSSS this idea correctly identifies the attractor on any given time slice. Let $\vec{X}(\tau) = \left(\tau_{0} \, T(\tau),\tau_0^2 \, \dot{T}(\tau) \right)$ be a point in a slice of phase space at time $\tau$ and $\tau_{0}$ denotes initialization time. This point moves with velocity $\vec{V}=\tau_{0} \dot{\vec{X}}(\tau)$. The slow region is defined by its Euclidean norm $V$, which has a minimum at asymptotically late times when 
	the system approaches local thermal equilibrium. However, the whole region where it is small is of dynamical significance. In Fig.~\ref{plot:descent}, the background color is determined by~$V$, where  bluer color implies lower speed. There is a slow region stretching out from local thermal equilibrium, and the attractor $\mathcal{A}_{*}(w)$ lies along it. It can be approximated by a slow-roll approximation~\cite{Heller:2015dha} where one neglects~$\ddot{T}$ in the equations of~motion. This generalizes directly to phase spaces of any dimension. For the case of HJSW the slow regions are shown also in blue in the bottom row of Fig.~\ref{plot:HJSW}. 
	
	In non-autonomous systems the slow region changes with time. If it were to evolve faster than the solutions do, it would not be useful to characterize the attractor. This is however not the case here: in both Fig.~\ref{plot:descent} and Fig.~\ref{plot:HJSW} we observe that once solutions reach the slow-region, they stay inside and evolve with it. This is analogous to the adiabatic evolution observed in the case of the Boltzmann equation in the relaxation time approximation considered in Ref.~\cite{Brewer:2019oha}.
	
	\mysection{Multidimensional phase spaces} In realistic settings, one may not be able to consider the full phase space, but only some finite-dimensional, approximate representation thereof. As an example of such a procedure, we have studied a $16$-dimensional subspace of the phase space of the Boltzmann kinetic equation in the relaxation time approximation~\cite{Baym:1984np, Florkowski:2013lya}. This subspace is captured by taking the lowest $16$ moments of the distribution function and solving the evolution equations as in Ref.~\cite{Strickland:2018ayk, Strickland:2019hff}. 
	We followed the evolution of a set of 240 initial conditions spanning a $5$-dimensional manifold embedded in the $16$-dimensional phase. 
	The results of an exploratory analysis of the resulting dataset using PCA are illustrated by Fig.~\ref{fig:RTA_EVR_tau} and described in more detail in the Supplementary Material \footnote{See Supplemental Material below for more details on dimensionality reduction in kinetic theory, which includes Refs.\,\cite{Heller:2016rtz, Heller:2018qvh}.}. This preliminary study supports the viability of the proposed approach in multidimensional phase spaces. 
	
	\begin{figure}[t]
		\centering
		\includegraphics[width=.75\columnwidth]{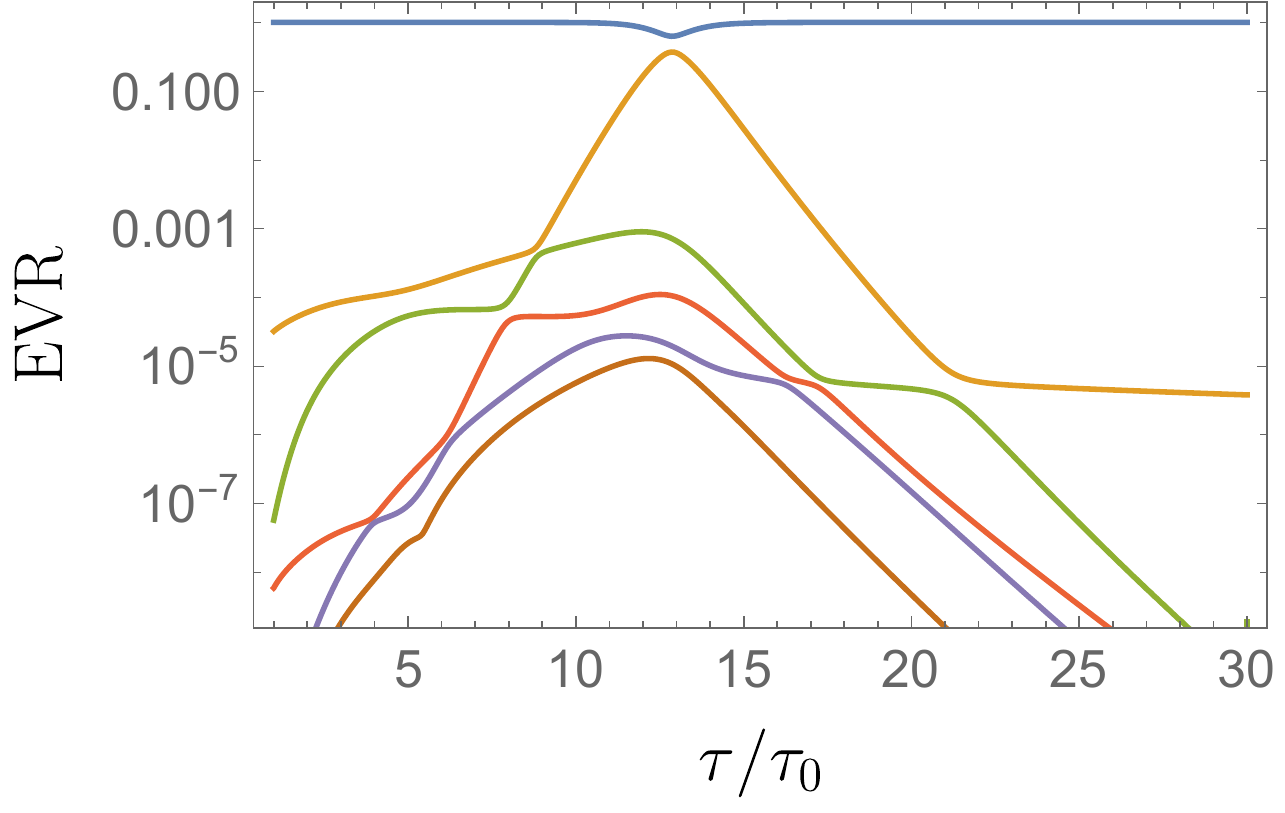}
		\caption{The six largest principal components and their explained variance ratios as a function of $\tau$. Due to the bias in initial conditions, one component dominates already at the initial time. This direction is not dynamically preferred,  as the other components initially grow in importance
			and only at $\tau \approx 15 \, \tau_{0}$ we see their expected exponential decay. At $\tau \approx  22 \, \tau_{0}$, the second principal component stabilizes. This feature 
			may reflect the limitations of PCA to capture 
			effects of curved regions in phase space.}
		\label{fig:RTA_EVR_tau}
	\end{figure}
	
	\mysection{Summary and outlook} The pressure anisotropy ${\cal A}(w)$ has been observed to exhibit universal behaviour in various models, characterized by different authors in terms of concepts such as 
	pullback/forward attractors or slow-roll, all of which capture some aspects information loss in dissipative systems.
	In this work we have described hydrodynamic attractors using concepts which allow for generalizations to more realistic settings.
	We have shown that this can be achieved by considering the phase space of the theory, which alleviates the need to know  
	in advance which particular quantity makes attractor behaviour manifest. From this perspective, the hydrodynamic attractor is associated with the more general notion of dimensionality reduction of phase space regions. 
	
	\mysection{Acknowledgements} We thank J.~Carrasquilla, R.~Janik, D.~Teaney as well as participants of the \emph{XIV Polish Workshop on Relativistic Heavy-Ion Collisions} at AGH, \emph{Initial Stages 2019} at Columbia University and \emph{Foundational Aspects of Relativistic Hydrodynamics} at Banff International Research Station where this work was presented at different stages of development for helpful discussions. The Gravity, Quantum Fields and Information group at AEI is supported by the Alexander von Humboldt Foundation and the Federal Ministry for Education and Research through the Sofja Kovalevskaja Award. MS was supported by the Polish National Science Centre grant 2018/29/B/ST2/02457.

	\bibliography{apsmain}
	\bibliographystyle{bibstyl}
	
	\balancecolsandclearpage
\begin{center}
	\textbf{\large Supplemental Material: Hydrodynamic attractors in phase space}
\end{center}
\setcounter{equation}{0}
\setcounter{figure}{0}
\setcounter{table}{0}
\setcounter{page}{1}
	\renewcommand{\theequation}{S\arabic{equation}}
	\renewcommand{\thefigure}{S\arabic{figure}}

	\section{Dimensionality reduction in Kinetic theory} 
	
	As a test of our approach in a somewhat more realistic example, we present some details of its application to the dynamics of kinetic theory, specifically the Boltzmann equation in the relaxation time approximation (RTA). This setting was examined in the context of attractors in a number of recent papers~\cite{Romatschke:2017vte, Romatschke:2017acs, Kurkela:2019set}. Most relevant for us is the work of Strickland~\cite{Strickland:2018ayk, Strickland:2019hff}, who considered the dynamics of the moments of the distribution function and demonstrated that they too show attractor behaviour. In this work we study the same collection of moments but instead of considering each of them separately, we view them as coordinates on a truncated phase space. From this perspective one apply the approach proposed in this Letter to identify correlations between moments and determine the effective dimensionality of a set of solutions as it evolves. 
	
	In RTA kinetic theory, in boost-invariant flow, the distribution function satisfies \cite{Baym:1984np, Florkowski:2013lya}
	\begin{multline}
	f(\tau,u,p_T) = D(\tau,\tau_0) f_0(u,p_T)  \\ + \int_{\tau_0}^\tau \frac{d\tau'}{\tau_{\text{eq}}(\tau')} D(\tau,\tau')f_{\text{eq}}(\tau',u,p_T),
	\end{multline}
	where $f_0$ defines the initial conditions. $u, \tau$ and $p_T$ are boost invariant variables and it is convenient to define $v \equiv \sqrt{u^2+p_T^2 \tau^2}$. The moments of the distribution function are defined by
	\begin{equation}
	M^{nm}[f] \equiv \int \frac{du}{v} d^2p_T \left(\frac{v}{\tau}\right)^n \left(\frac{u}{\tau}\right)^m f.
	\end{equation}
	We follow Ref.~\cite{Strickland:2018ayk, Strickland:2019hff} closely in solving for the evolution of moments. Each moment satisfies
	\begin{multline}
	M^{nm}(\tau) = D(\tau,\tau_0) M^{nm}[f_0](\tau)  \\ + \int_{\tau_0}^\tau \frac{d\tau'}{\tau_{\text{eq}}(\tau')} D(\tau,\tau')M^{nm}[f_\text{eq}](\tau')~.
	\end{multline}
	For an equilibrium distribution of the Boltzmann form
	\begin{equation}
	f_\text{eq}(\tau,u,p_T) = e^{-\frac{v}{T(\tau) \tau}},
	\end{equation}
	each $M^{nm}[f_\text{eq}]$ can be calculated analytically and are determined by $T(\tau)$. The temperature is related to the energy density by Landau matching and it follows that the equation for $M^{20}$ translates into a closed equation for the temperature   
	\begin{multline}
	T^4(\tau) = D(\tau,\tau_0) T^4[f_0](\tau) \\ + \int_{\tau_0}^\tau \frac{d\tau'}{2\tau_{\text{eq}}(\tau')} D(\tau,\tau')T^4(\tau') H\left(\frac{\tau'}{\tau}\right),
	\end{multline}
	where $H(y) = y^2+\frac{\tan ^{-1}\left(\sqrt{\frac{1}{y^2}-1}\right)}{\sqrt{\frac{1}{y^2}-1}}$. Solving this equation for $T(\tau)$ allows one to calculate every moment $M^{nm}(\tau)$. As in the main text of our Letter, we consider a conformal theory with 
	\begin{equation}
	T \tau_\text{eq} = 5 \eta/s
	\end{equation}
	and we take $\eta/s = 1/(4 \pi)$. The scaled moments
	\begin{equation}
	\label{eq.scaledmoments}
	\bar{M}^{nm}(\tau) = M^{nm}(\tau)/M^{nm}_{\text{eq}}(\tau),
	\end{equation}
	which all tend to unity at equilibrium, provide a natural setting in which to study phase space dimensionality reduction in this model. This space is infinite-dimensional but as in Refs.~\cite{Strickland:2018ayk, Strickland:2019hff} we direct our attention to the 16 moments with $n$ and $m \leq 3$.
	
	At large $\tau$ the effective temperature evolves according to the Bjorken asymptotics (see Eq.~\eqref{bjorken}) up to exponentially decaying corrections which reflect the spectrum of non-hydrodynamic modes \cite{Heller:2016rtz, Heller:2018qvh}. Given the time evolution of individual moments~\cite{Strickland:2018ayk, Strickland:2019hff}, within the phase space picture we expect that a single principal component dominates the late time signal, while the rest should decay exponentially. 
	
	\begin{figure*}[t]
		\centering
		\includegraphics[width=.75\textwidth]{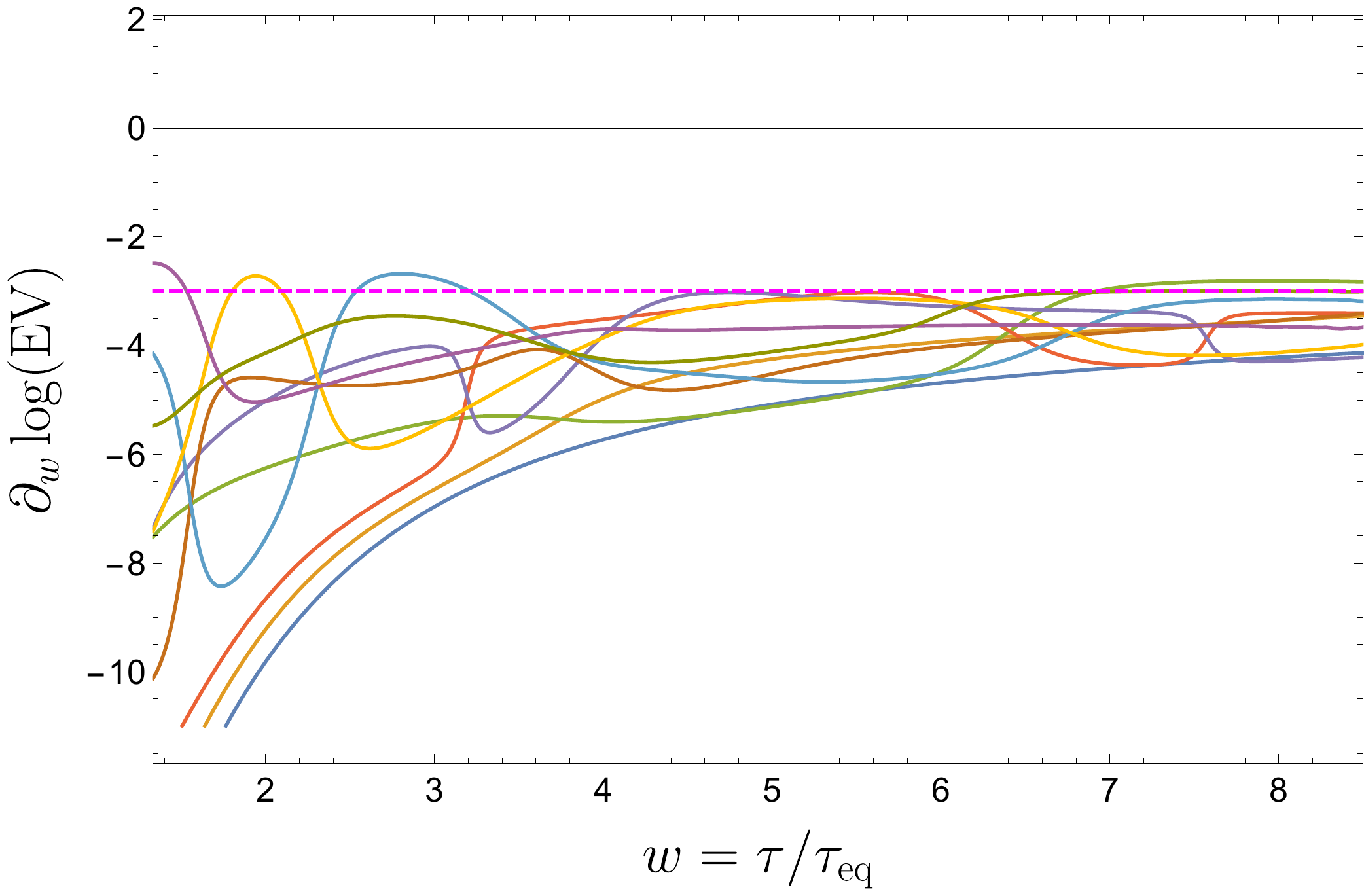}
		\caption{The derivative of the logarithm of the explained variances seems to approach a constant at large $w$, indicating an exponential decay. All the components have a similar decay rate, showing that the collapse is somewhat uniform. As was observed in BRSSS (see Fig.~\ref{fig:brsss_pca} in the main text), this decay rate is consistent with twice the decay rate of the non-hydrodynamic modes (dashed magenta line). The oscillations whose frequency decreases at later times is also a feature shared by the non-hydrodynamic modes, see Eq.~\eqref{eq:rta-nonhydro}.}
		\label{fig:RTA_EV_decay}
	\end{figure*}
	
	Defining the dimensionless time variable $w \equiv \tau/\tau_{\text{eq}}$ (the normalization of this variable differs by a constant factor from what is used in the main text), the behaviour is even simpler. The power law piece of different solutions becomes universal and the difference between two solutions at late times is to leading order
	\begin{equation}\label{eq:rta-nonhydro}
	\delta M^{nm} \sim e^{-\frac{3}{2}w} w^\beta,
	\end{equation}
	where $\beta$ is a certain set of complex numbers \cite{Heller:2018qvh}. Thus, applying PCA to the set of solutions as a function of $w$, the expectation is that all of them decay exponentially with the same rate asymptotically. Note also that the imaginary part of $\beta$ leads to oscillations in logarithmic time.
	
	We choose as our family of initial conditions
	\begin{equation}
	f_0(u,p_T) = A(T_0) e^{-\frac{\sqrt{u^2/\alpha_0^2 + p_T^2\tau_0^2}}{\Lambda_0 \tau_0}}  \lvert u \rvert ^{c_u} \lvert p_T \rvert^{c_p},
	\end{equation}
	where $A(T_0)$ is a normalization factor ensuring that the initial temperature is $T_0$. These initial conditions depend on five parameters $ (T_0, \Lambda_0, \alpha_0, c_u, c_p)$. For these initial conditions, it is easy to evaluate $M^{nm}[f_0]$ to arbitrary precision. This is the reason why they were used earlier in Ref.~\cite{Heller:2018qvh} to explore hydrodynamization patterns in kinetic theory. Sampling these five parameters 
	at the initial point in time leads to an, at most, five-dimensional manifold embedded in a 16-dimensional space. We sample uniformly in (0.8-1.2 GeV, 0.8-1.2 GeV, 0.5-1.4, -0.5-1, -0.5-1) and take $\tau_0 = 0.1$ fm/c.

	In the BRSSS and HJSW models to which most of the main text was devoted, we had much more control over initial conditions and could choose them without bias in any particular direction. The explained variances were initially equal and the hierarchy that developed was purely a result of the dynamics. The present case is more challenging, as the initial conditions 
	we use have a built in bias which translates into a hierarchy of explained variances already at the initial time. Nevertheless, the evolution exhibits the expected characteristics of dimensionality reduction also in this restricted setting. In Fig.~4 in the main text, we plot the explained variance ratio of the six largest principal components as a function of $\tau$. At early times, the bias in initial conditions is the reason for the large difference in explained variance. As time elapses, the relative importance of the smaller ones grow quickly, showing that the initial situation was not dynamically preferred. At $\tau \approx 15 \, \tau_{0}$, dimensionality reduction occurs leaving a single dominant component while the rest decays exponentially. Surprisingly, at $\tau \approx 22 \, \tau_{0}$, the second component stabilizes, albeit at a very small value. We conjecture that this is due to the inherent limitations of PCA to capture curved manifolds. If the curve corresponding to the hydrodynamic modes has curvature, more principal components will be needed to describe it, resulting in the behaviour we see in the figure. It would be interesting to apply more sophisticated data analysis methods that can capture such non-linearity, e.g. kernel-PCA.
	
	To characterize the decay rate, we use PCA on the solutions as a function of $w$. In Fig.~\ref{fig:RTA_EV_decay} we plot the derivative of the logarithm of the explained variances. At large $w$, they seem to saturate to a constant which is consistent with the twice the decay rate of the non-hydrodynamic modes. In addition, at earlier times we see oscillations, whose frequency seem to decrease at later times. This strengthens the connection to the non-hydrodynamic modes, which exhibit oscillations in logarithmic time.
	
	We have demonstrated that PCA can be used to study the dynamics of dimensionality reduction in theories with higher dimensional phase spaces. While kinetic theory is in fact infinite-dimensional, all of this information will never be physically relevant and one may truncate the to a finite dimensional space (in this case, the lower moments of the distribution function). At late times, when the dimensionality reduction is governed by interactions, we can interpret the evolution of explained variances in terms of hydrodynamic and non-hydrodynamic modes. At earlier times, when the expansion is dominant, the character of the collapse will be different \cite{Kurkela:2019set}. We leave this analysis to future work.

	
\end{document}